\documentclass[%
 reprint,
 superscriptaddress,
 amsmath,amssymb,
 aps,
prl,
]{revtex4-1}

\usepackage{graphicx}%
\graphicspath{ {Figures/} }
\usepackage{dcolumn}%
\usepackage{bm}%
\usepackage{xcolor}
\usepackage{mhchem}
\def\af{$\alpha$-Fe}

\begin{document}

\title{
Hydrogen Diffusion and Trapping in $\alpha$-Iron:\\ The Role of Quantum and Anharmonic Fluctuations}%

\author{Bingqing Cheng}
\affiliation{
Laboratory of Computational Science and Modeling, Institute of Materials,
    \'Ecole Polytechnique F\'ed\'erale de Lausanne, 1015 Lausanne, Switzerland
}

\author{Anthony T. Paxton}
 
\affiliation{Department of Physics, King's College London. Strand, London WC2R 2LS, UK
}

\author{Michele Ceriotti}
\affiliation{
Laboratory of Computational Science and Modeling, Institute of Materials,
    \'Ecole Polytechnique F\'ed\'erale de Lausanne, 1015 Lausanne, Switzerland
}

\date{\today}%

\begin{abstract}
We investigate the thermodynamics and kinetics of a hydrogen interstitial in magnetic $\alpha$-iron, taking account of the quantum fluctuations of the proton as well as the anharmonicities of lattice vibrations and hydrogen hopping.
We show that the diffusivity of hydrogen in the lattice of BCC iron deviates strongly from an Arrhenius behavior at and below room temperature. 
We compare a quantum transition state theory to explicit ring polymer molecular dynamics in the calculation of diffusivity and we find that the role of phonons is to inhibit, not to enhance, diffusivity at intermediate temperatures in constrast to the usual polaron picture of hopping. 
We then address the trapping of hydrogen by a vacancy as a prototype lattice defect. By a sequence of steps in a thought experiment, each involving a thermodynamic integration, we are able to separate out the binding free energy of a proton to a defect into harmonic and anharmonic, and classical and quantum contributions. 
We find that about 30\% of a typical binding free energy of hydrogen to a lattice defect in iron is accounted for by finite temperature effects and about half of these arise from quantum proton fluctuations. This has huge implications for the comparison between thermal desorption and permeation experiments and standard electronic structure theory. The implications are even greater for the interpretation of muon spin resonance experiments.
\end{abstract}

\maketitle

The injection, transport and trapping of subatomic particles such as protons, deuterons, tritons, muons, or positrons in solids takes a pivotal role in experimental characterization techniques such as muon spin spectroscopy ($\mu$SR)~\cite{Brandt1986}, positron annihilation experiments~\cite{Seeger1974,Slotte2016}; and in the design of plasma containment in fusion power generators~\cite{Causey2001}.
In the case of hydrogen, diffusion and trapping is also crucial in many technological and materials science applications, including for instance hydrogen storage and fuel cells~\cite{Cabot2000,Alique2018}, in particular the deleterious effects of hydrogen on electrode integrity as a consequence of the Gorski effect. 
Diffusion of hydrogen in iron is also of interest in final stages of stellar evolution~\cite{VanHorn1970}.
The problem of hydrogen embrittlement of iron and steel is deeply concerned with the rate of proton diffusion and the depth of lattice defect traps which may serve to attenuate the diffusivity, since it is expected that crack tip speed may be limited by the rate at which it can be fed by hydrogen~\cite{Barrera2018}. 
The trapping of hydrogen by vacancies is of particular importance because by the defactant effect~\cite{Kirchheim2007} the vacancy is stabilized by trapping and indeed the equilibrium vacancy concentration is known to be enhanced by orders of magnitude as a result of hydrogen ingress~\cite{Fukai2005} leading to damage and compromised structural integrity~\cite{Takai2008}. The depth of a microstructural trap, that is, the free energy gain by transferring a proton from a bulk tetrahedral site into the trap, is extremely hard to measure since although an average trap depth over many defects is accessible through thermal desorption spectroscopy, it is not possible to prepare specimens with just a single defect in order to distinguish, say a dislocation trap from a grain boundary or interface site. This is particularly difficult in the case of the vacancy. It is possible to calculate trap depths using density functional theory (DFT) and for example it has been shown that the vacancy may trap up to six protons---one close to each face of the cube surrounding the defect~\cite{tateyama2003stability}. Unfortunately standard DFT calculations have ready access only to the zero temperature total energy the quantum nature of the subatomic particle is usually neglected or accounted for only in terms of a zero point energy (ZPE).
Our aim in the present work is to unravel various contributions to the binding free energy to provide both a framework for the general case and to address the trapping of H in ferrite ($\alpha$-Fe) quantitatively.
Atomistic modelling of a hydrogen interstitial in $\alpha$-iron poses enormous challenges; first
the magnetism requires an explicit treatment of the electronic degrees of freedom of the system~\cite{tateyama2003stability,counts2010first,paxton2014quantum},
second the timescale required to measure the H binding free energy to defects as well as the diffusivity in the bulk lattice 
is usually not accessible in ab initio MD simulations.
To make the matter even more complex, the small mass of the proton means that nuclear quantum effects (NQEs) can play an important role at room temperature and below. 
For instance, it has been demonstrated that NQEs have a significant effect on the thermodynamic stability of different phases~\cite{mora+13prl,engel2015anharmonic,rossi2016anharmonic}, as well as on the diffusivity of protons~\cite{kimi+11prb,hass+13pnas,ross+16jpcl}.

In the present study interatomic forces are described within the self consistent magnetic tight
binding (TB) approximation~\cite{Paxton2008}. 
Parameters for the model are
given in Ref.~[\onlinecite{Paxton2013}].
TB theory is an abstraction of the DFT and hence has the benefit of capturing the essential physics of the chemical bond, including self consistent charge transfer, with forces derived from the Hellmann-Feynman theorem, but because the Hamiltonian is obtained from a fitted look-up table rather than determined ab initio, the method is computationally very fast.
The evaluation of NQEs
is achieved by using the imaginary time path
integral formalism of quantum mechanics. 
The path integral formalism
maps the quantum mechanical partition function onto the
partition function of a classical ring-polymer system~\cite{feyn-hibb65book,chan-woly81jcp,parr-rahm84jcp,ceri+10jcp},
and as such the quantum system can be described by $P$ copies of the physical system
with corresponding particles in adjacent replicas connected by harmonic springs.
When $P=1$ the nuclei are purely classical, and when $P \rightarrow \infty$ each nucleus in the ring polymer system is fully consistent with the statistics of a quantum system of distinguishable particles.
Methods inspired by path integral molecular dynamics (PIMD)~\cite{cao-voth94jcp,crai-mano04jcp} can also be used to approximate time-dependent observables. We will use the thermostatted ring polymer molecular dynamics (TRPMD) method~\cite{ross+14jcp}. 
The reader is referred to recent reviews for a more thorough discussion of PIMD-related methods~\cite{habe+13arpc}.%

\begin{figure}
    \centering
    \includegraphics[width=0.5\textwidth]{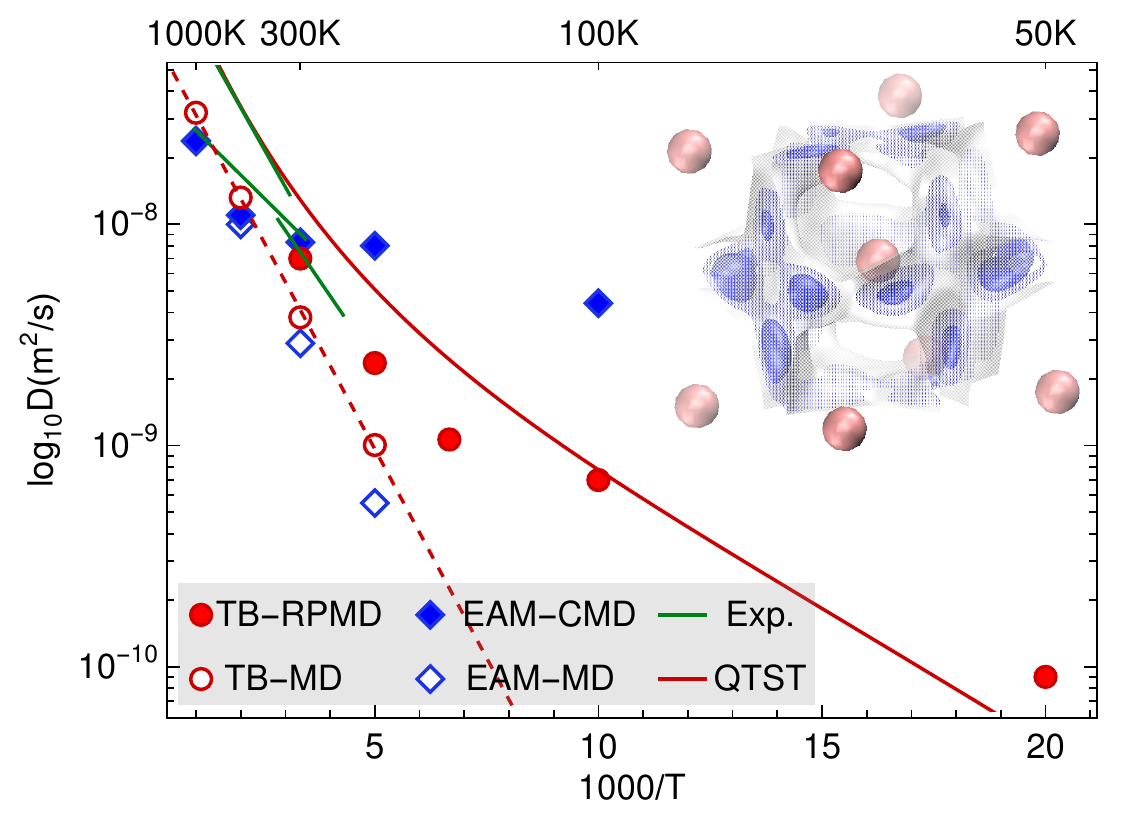}
    \caption{(Color online) Measured and calculated diffusivity of hydrogen in \af. Experimental data are drawn as solid green Arrhenius lines in the temperature range of the measurements~\cite{Nagano1982,Kiuchi1983}. Blue squares are from centroid molecular dynamics simulations by Kimizuka {\it et al.}~\cite{Kimizuka2011} using an embedded atom classical potential (EAM) and a Morse potential for hydrogen and iron; MD is classical molecular dynamics and CMD is centroid MD.  The red curve is the result of a quantum transition state theory (QTST) calculation by Katzarov and Paxton~\cite{Katzarov2014,Paxton2016}.
    The open and solid red squares are our results using the TB hamiltonian and classical MD and path integral MD (RPMD), respectively.
    On the inset, the distribution of quantum mechanical H atoms in the BCC unit cell that was computed from RPMD simulations is shown. 
    }
    \label{fig_Diff-H}
\end{figure}

In order to compute the quantum configurational distribution and the
diffusivity of H in bulk \af\ lattice,
we first performed TRPMD simulations of a system consisting of 
16 Fe atoms on a perfect BCC iron lattice, and a H interstitial atom.
We performed simulations at 300, 200, 150, 100 and 50 K, increasing the number of beads $P$ from 16 to 64 beads as the temperature was lowered, to account for the stronger quantum nature of nuclei at lower temperature. 
For the sake of comparison,  we also performed classical simulations (i.e. only using one bead for the ring polymer) from 1000 K to 100 K. 
The diffusion coefficients of H in the bulk \af\ lattice were computed 
from the $\omega\rightarrow 0$ limit of the velocity-velocity autocorrelation spectrum of the H atom.
Results for these simulations are reported in Figure~\ref{fig_Diff-H}, compared with the 
results from a previous calculation using a classical embedded atom potential (EAM) for interatomic forces~\cite{Kimizuka2011},
as well as experimental measurements~\cite{Nagano1982,Kiuchi1983} in the high-temperature regime.

The most prominent observation from Figure~\ref{fig_Diff-H} is the stark difference between the classical and the quantum diffusion coefficient of H in the bulk lattice at temperatures equal or lower than room temperature.
Using classical molecular dynamics, the temperature dependence closely follows Arrhenius behavior as indicated by the dashed red line.
Furthermore, the classical diffusivities predicted by the TB hamiltonian and the EAM force field are very similar, despite the distinct forms of the potentials.
However, when nuclear quantum effects are included by either using the path integral formalism or by employing 
centroid molecular dynamics,
a strong deviation of the H diffusivity from Arrhenius behavior below room temperature emerges.
At the low-$T$ end of the experimental temperature range (300~K), classical MD predictions for both TB and EAM are about a factor of two lower than experiments, while quantum results for both models are in good agreement. 
The discrepancy between classical and quantum dynamics indicates the importance of NQEs, which becomes dramatic  at lower temperatures (50~K to 200~K).

While the EAM and TB are in agreement in the classical MD, there is a large discrepancy at low $T$ between the EAM-CMD and the TB-RPMD. 
It is not unusual to see larger discrepancies between potential energy surfaces when simulations are performed that include nuclear quantum fluctuations, because configurations explore regions that display large levels of anharmonicities, and that are often not included in the fitting of the potential~\cite{Malerba2010,wang+14jcp}. 
It is possible that while the EAM is fitted to the classical activation barrier, the EAM does not describe well the three dimensional potential energy surface for H moving around the Fe-lattice. On the other hand the TB very well reproduces this ``adiabatic surface'' in comparison to density functional calculations, patricularly near the saddle point~\cite{Paxton2010}. This means that as the beads wander far from the classical reaction coordinate the proton samples regions of the configuration space that the EAM does not describe well.

It is very instructive to compare the QTST with the TRPMD. The QTST uses fixed potential energy surfaces at the reactant basin and at the saddle point. After that the partition functions are calculated, which means that the rate coefficient can be found without great effort at any temperature~\cite{Paxton2016}. 
Because the rather artificial construction is made in Ref.~\citenum{Paxton2016} that the potential energy in configuration space is calculated using a relaxed atomic system with the proton constrained at the saddle point the QTST would be expected to overestimate the diffusivity. 
On the other hand, the QTST neglects dynamic phonon effects such as phonon assisted tunneling in the polaron picture~\cite{FlynnStoneham}. 
From that point of view one might expect that the TB-RPMD would predict a greater diffusivity than the TB-QTST. As seen in figure~\ref{fig_Diff-H} the opposite is the case at temperatures between 100 and 300~K.
This implies that phonon assisted tunneling is not a large effect: on the contrary our calculations reveal that phonons attenuate proton diffusion, presumably through proton-phonon scattering.
The qualitative agreement between TB-QTST and TB-RPMD at all temperatures validates the use of the much cheaper QTST to estimate rate coefficients in the quantum regime. 

To elucidate the mechanism of H diffusion in \af\, 
we show the quantum mechanical density distribution of H at 300~K in a BCC unit cell on the inset of Figure~\ref{fig_Diff-H}.
It can be seen that the equilibrium positions for H in the lattice are tetrahedral (T) sites.
Meanwhile, NQEs broaden the spread of the distribution of H around the energy minima, 
which indicates strong ZPE effects in the H hopping.
In other words, NQEs delocalize H in the reactant state, and effectively reduce the free energy barrier for H migration between neighboring T-sites.

Diffusion in the perfect BCC lattice is a necessary component of the mechanistic understanding of the mobility of H in \af. However, the rate-limiting step for macroscopic diffusion always involves binding to crystal defects. 
In order to assess the importance of different terms in the overall binding free energy between the H atom and the defect,
in the second part of our study we consider the archetypical example of a vacancy in \af. 
Computing this binding energy by sampling of the NVT ensemble is difficult, as the waiting time for a trapped H to be released is far beyond the time scale of standard molecular dynamics. 
Furthermore, a very large simulation would be needed to bring the H atom sufficiently far from the vacancy to estimate accurately the binding energy in the dilute limit. 

\begin{figure}
    \centering
    \includegraphics[width=0.45\textwidth]{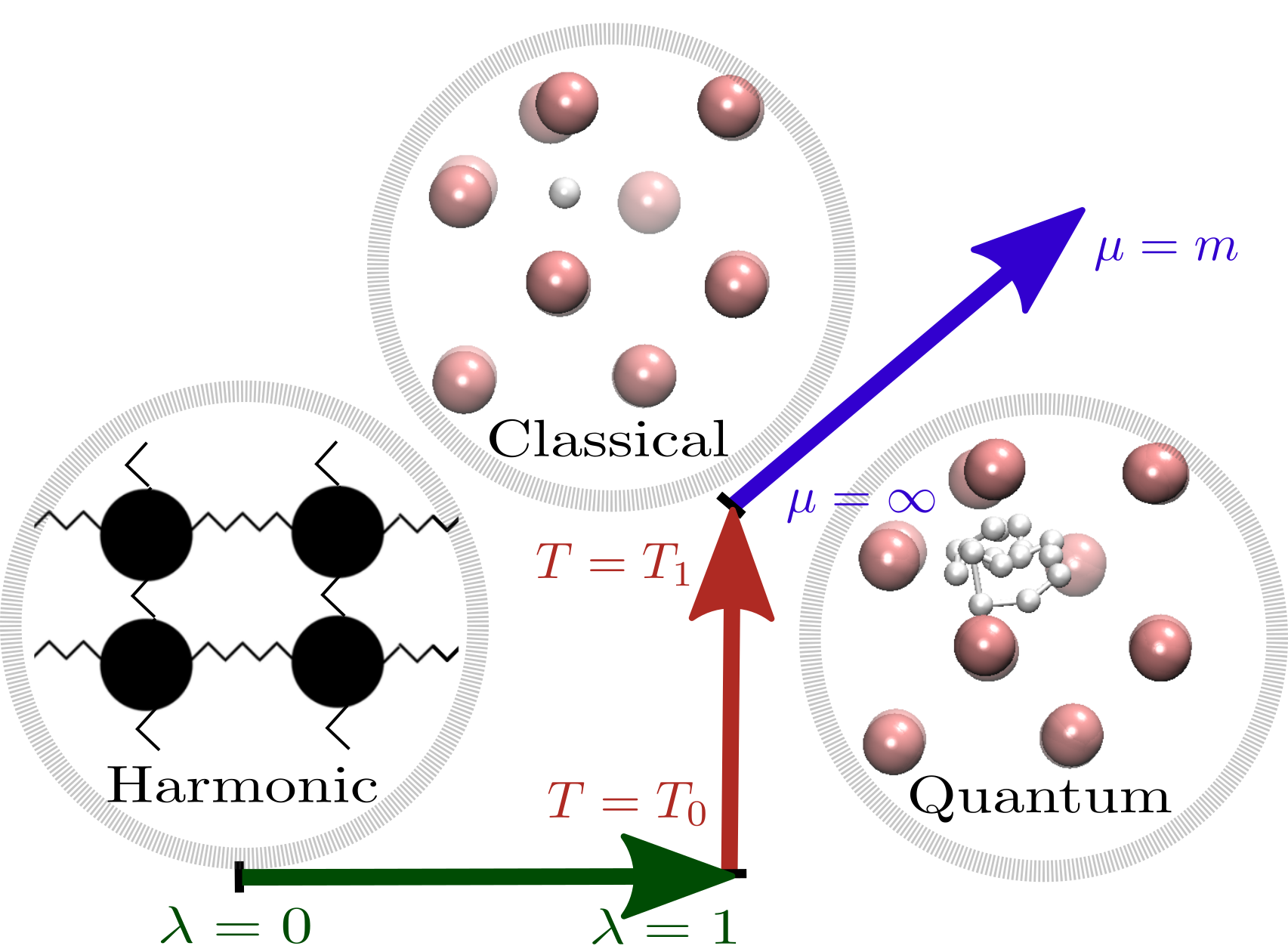}
    \caption{(Color online) Schematics of thermodynamic integration (TI) routes used in the free energy evaluations. 
    The green arrow indicates the switching between an harmonic reference system ($\lambda=0$) and a real system ($\lambda=1$),
    the red arrow illustrates TI with respect to temperature,
    and the blue arrow shows TI from a classical to a quantum mechanical system. }
    \label{fig:tdi}
\end{figure}

For these reasons, we decided to compute the stability of a H atom bound to a vacancy, relative to that of a H atom in a tetrahedral site of the perfect BCC lattice, by computing first the absolute Helmholtz free energies for four systems separately~\cite{footnote}:
({\it i\/}) a perfect bulk \af\ system that has 16 atoms (\ce{Fe16}),
({\it ii\/}) a system with a vacancy (\ce{Fe15}),
({\it iii\/}) a system with a H interstitial (\ce{Fe16H}), and 
({\it iv}) a system with a vacancy and a H interstitial (\ce{Fe15H}).
Based on the Helmholtz free energy of the four independent systems,
at a certain thermodynamic condition the binding energy of a H to a mono-vacancy can be schematically expressed as $A_{V-H} = A(\ce{Fe16H}) + A(\ce{Fe15}) - A(\ce{Fe15H}) - A(\ce{Fe16})$.

To compute $A$ for the four systems, we used the thermodynamic integration (TI) method, that uses a series of simulations of real or artificial systems to compute the various components of the free energy difference between a harmonic reference system and the fully anharmonic, quantum system. 
To do so efficiently, 
we have carefully selected a combination of multiple thermodynamic integration routes
as depicted schematically in Figure~\ref{fig:tdi}.
This combination thus takes into account vibrational entropy, anharmonicity and NQEs, and makes it possible to disentangle the different contributions.
Since a detailed description of thermodynamic integration routes and several tricks of the trade
can be found in Ref.~\cite{cheng2017computing},
here we briefly describe the routes employed in the present study.
The first TI route (the green arrow in Figure~\ref{fig:tdi}) goes from the classical harmonic crystal whose free energy $A_{\text{h}}$ is analytic, to the classical physical system at a low temperature $T_0=$10 K,
under which condition the H interstitial atom does not jump between degenerate trap sites about the vacancy during the MD simulations.
The second TI route (the red arrow) allows us to obtain the temperature dependence of the Helmholtz free energies of each Fe-H system,
by running simulations of the classical physical system under the NVT ensemble
from the low temperature $T_0$ to a higher temperature $T_1=300$ K.
The last part of TI takes into account  NQEs at $T_1=300$ K.
The quantum mechanical free energy difference is a function of atomic mass, 
thus for a given system the overall NQEs in free energy can be evaluated from the integration of the quantum centroid virial kinetic energy with respect to the fictitious “atomic” mass $\mu$~\cite{fang2016inverse,rossi2016anharmonic}. 
In practice, the integrand was evaluated for the actual system and for systems with all the atomic masses scaled 4 and 16 times 
in PIMD simulations.

\begin{figure}
    \centering
    \includegraphics[width=0.45\textwidth]{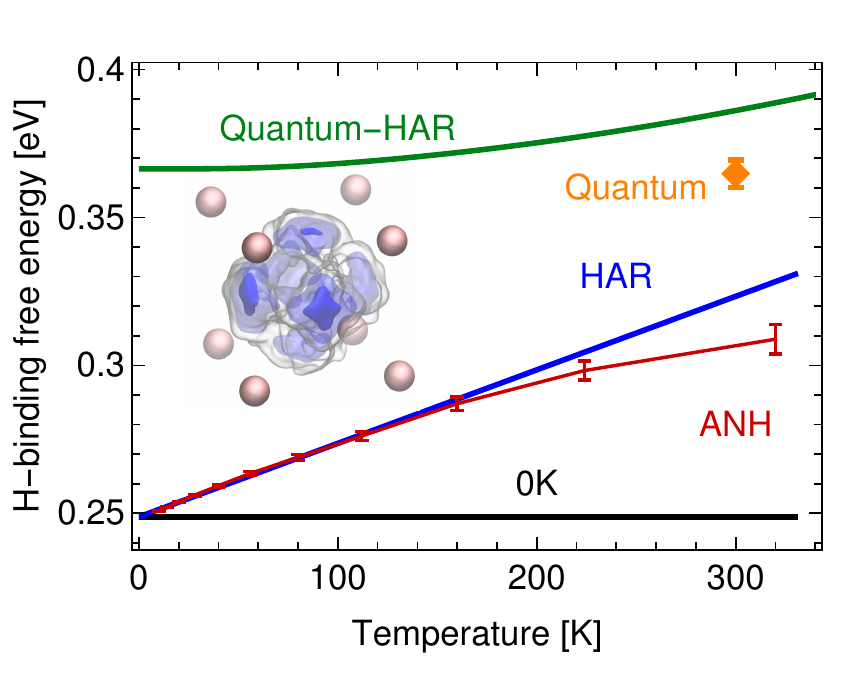}
    \caption{The temperature dependent H binding free energy to a mono-vacancy in \af\ .
    The black line is the prediction just using the minima of the potential energy surface at 0 K,
    the blue line shows the harmonic approximation for the classical system,
    the green line illustrate the harmonic approximation for the quantum mechanical system,
    the red curve indicates the fully anharmonic result of the classical system,
    and the yellow dot shows the quantum and anharmonic result.
Statistical uncertainties are indicated by the errorbars.
In the inset: the distribution of quantum mechanical H atoms near a vacancy, as computed by TRPMD at 300~K. The proton spends no time at the vacant site itself; this is consistent with DFT calculations~\cite{tateyama2003stability} and validates our TB hamiltonian.}
    \label{fig_avh}
\end{figure}

In Figure~\ref{fig_avh} we plot
the predictions from the harmonic approximations,
the classical anharmonic free energy contribution,
and the overall free energy of binding taking into account fully both anharmonicity and NQEs.
At all temperatures, the vibrational entropy plays an important role in the hydrogen-vacancy binding energy of the classical Fe-H systems,
as demonstrated by the considerable difference between the 0 K prediction (the black line in Figure~\ref{fig_avh}) and classical harmonic approximation (the blue line in Figure~\ref{fig_avh}).
This difference in the vibration frequencies of the vacancy-trapped and the free H in the T sites also translates to the large zero point energy contribution to the overall binding free energy,
which is indicated by the remarkable gap between the harmonic approximations using the classical Boltzmann distribution and the quantum mechanical Bose-Einstein  distribution (the blue line and the green line in Figure~\ref{fig_avh}, respectively).
Finally, anharmonicity, which is an oft-neglected part in previous DFT calculations~\cite{tateyama2003stability,vekilova2009first,lu2005hydrogen,johnson2010hydrogen,counts2010first,ohsawa2012configuration}
has a non-negligible effect even at room temperature.
The anharmonic part of the potential is also softer for the H in the vacancy than for the interstitial site, and so the anharmonic contribution to the free energy leads to a further stabilization of the bound state, both classically and quantum mechanically. 
In short, the anharmonic quantum mechanical trapping energy of H in alpha-iron at 300 K is predicted to be $0.365 \pm 0.005$ eV using the TB Hamiltonian.
This result compared well with the experimental measurements of $0.55-0.81$ eV for hydrogen trapping energy in alpha-iron ~\cite{kim1985vacancies,bhadeshia2016prevention}, and $0.48-0.63$ eV for deuterium at room temperature~\cite{myers1979defect,besenbacher1987multiple}, whereas a result based purely on the potential energy difference between the two states would predict a much lower binding energy of 0.25 eV. This has implications for the identification of trap sites by comparison of thermal desorption spectra and total energy calculations. Figure~\ref{fig_avh} shows that the usual correction of adding the ZPE~\cite{counts2010first} is in this case accurate since the temperature correction to the harmonic approximation at 300~K is canceled by the anharmonic correction---but this will not always be the case.

In conclusion, we have characterized the importance of NQEs and anharmonicity in two of the microscopic mechanisms that underlie the transport of H atoms in in \af\,
namely H diffusion in the perfect BCC lattice,
and the thermodynamic free energy that describes the binding of H to a monovacancy.
Nuclear quantum effects change the diffusion coefficient of H in bulk \af\ by a factor of two at room temperature, and the quantum effects become overwhelming at lower temperatures. 
We then consider the case of the binding free energy of H to a monovacancy, for which we considered and disentangled different contributions such as vibrational entropy, anharmonicity and NQEs, concluding that they all play a significant role at room temperature, and collectively increase the binding energy from 0.25 meV to 0.36 meV. 
This latter is closer to the experimental estimates, and the magnitude of the quantum contribution is consistent with the experimental observation that deuterium is less strongly bound than $^1$H. 
Our findings thus suggest that nuclear quantum effects may have significant effects on the interactions between H atoms and other defects, 
which are essential in achieving a quantitative predictive capability of the hydrogen embrittlement process. 
In addition, hopping and trapping of other charged subatomic particles in metal lattices is of central importance in solid state physics, encompassing phenomena such as $\mu$SR and positron annihilation experiments. 
There is no doubt in view of our findings that quantum fluctuations will take a greater part in the physics of these processes and this study has furnished us with a recipe for how to address these questions, such as the diffusivity of a positron in a metal or semiconductor or the trap depth of a muon at a crystal defect.

\begin{acknowledgements}
M.C. and B.C. acknowledge financial support by the Swiss National Science Foundation (project ID 200021-159896), and allocation of CPU time under the CSCS Project ID s787. ATP acknowledges the support of the UK EPSRC under the Programme Grant HEmS, EP/L014742.
\end{acknowledgements}

\end{document}